\documentclass[twocolumn,preprintnumbers,amsmath,amssymb]{revtex4}
\usepackage{tabularx,graphicx}

\usepackage{color}
\usepackage{hyperref}
\hypersetup{
    colorlinks=true,
    linkcolor=blue,
    filecolor=blue,      
    urlcolor=blue,
}

\usepackage{color}

\newcommand{\bluee}{\textcolor{black}}

\usepackage{ulem}   

\begin{document}

\newcommand{\beq}{\begin{equation}}
\newcommand{\eeq}{\end{equation}}
\newcommand{\beqn}{\begin{eqnarray}}
\newcommand{\eeqn}{\end{eqnarray}}
\newcommand{\bmath}{\begin{subequations}}
\newcommand{\emath}{\end{subequations}}
\newcommand{\bra}[1]{\langle #1|}
\newcommand{\ket}[1]{|#1\rangle}

\title{Does the Meissner effect violate the second law of thermodynamics? Comment on ``The Law of Entropy Increase and the Meissner Effect'' by A. Nikulov}

\author{J. E. Hirsch}
\address{Department of Physics, University of California, San Diego,
La Jolla, CA 92093-0319}
 
 \begin{abstract} 
In Entropy 24, 83 (2022) \cite{nikulov}, titled ``The Law of Entropy Increase and the Meissner Effect'', its author Alexey Nikulov argues  that the Meissner effect exhibited by
type I superconductors violates the second law of thermodynamics. 
Contrary to this claim, I show that the Meissner effect  is consistent with the second law of thermodynamics
provided that a mechanism exists for the supercurrent to start and stop without generation of Joule heat.
The theory of hole superconductivity provides such a mechanism, the conventional theory of superconductivity does not. It requires the existence of hole carriers in the normal state
of the system.
 \end{abstract}
 \maketitle

\section{Introduction}
Contrary to the overwhelming consensus in the physics community, A. Nikulov correctly points out  in Ref. \cite{nikulov} that the Meissner effect poses fundamental questions that are not
addressed in the conventional understanding of superconductivity. Nikulov  concludes from his analysis that the
Meissner effect is inconsistent with the second law of thermodynamics \cite{nikulov}. Because the Meissner effect is observed in nature, Nikulov concludes that 
this implies that the second law of thermodynamics is not valid for all natural processes, as generally assumed, and that
a perpetuum mobile of the second kind is realized by superconductors \cite{nikulov,perpetuum}.

Instead, I show in this paper that Nikulov's drastic conclusion can be avoided: the Meissner effect is shown to be consistent with the second law of thermodynamics, provided we make certain assumptions that are  inconsistent with the conventional BCS theory of superconductivity  \cite{tinkham}. These assumptions are instead consistent with
the alternative theory of hole superconductivity \cite{book,holesc,holescreview}.

       \begin{figure} [t]
 \resizebox{8.5cm}{!}{\includegraphics[width=6cm]{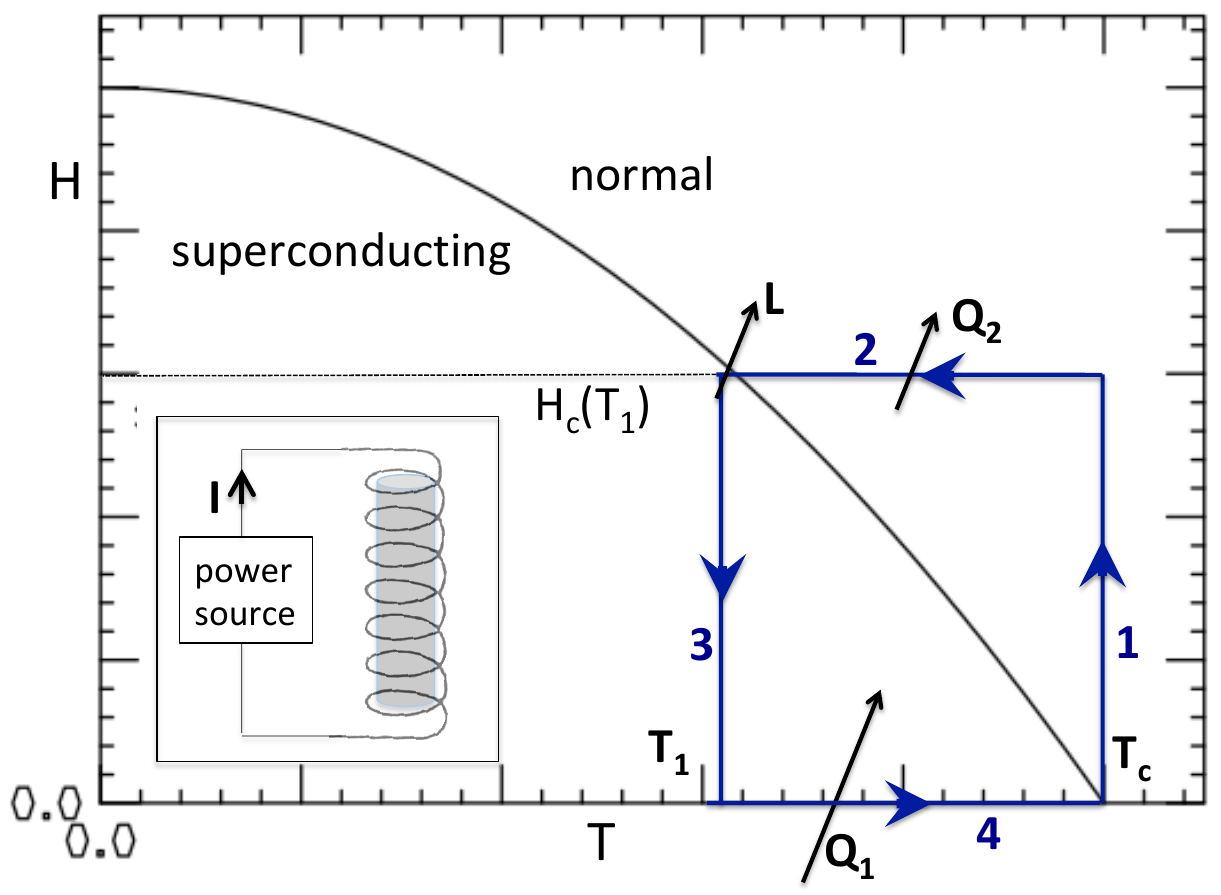}} 
 \caption { Phase diagram of a type I superconductor in the H-T plane. The transition is first order for finite field.
The blue lines show the 4 steps in the Gorter cycle labeled 1, 2, 3, 4. The inset shows the superconducting
sample (long cylinder) inside a solenoid through which current supplied by the power source flows, that
generates the magnetic field..}
 \label{figure1}
 \end{figure}

\section{The Gorter cycle}

Figure 1 shows a Gorter cycle for a long superconducting cylinder inside a solenoid powered by a power source.
The cycle starts at the critical temperature $T_c$ in the absence of applied field: $T_c \equiv T_c(H=0)$.
During step 1, the field is increased to value $H_c(T_1)$. During step 2, the system is cooled
in the presence of the magnetic field $H_c(T_1)$ from $T_c$ to infinitesimally below $T_1$. As the temperature is lowered from right above $T_1$ to right below $T_1$ the system
expels the magnetic field and enters the superconducting state (Meissner effect). 
During step 3, the magnetic field is lowered back to 0 at temperature $T_1$. During step 4, the system in the 
superconducting state is heated up to right above the critical temperature $T_c$ and enters the normal state.

The system releases heat $Q_2$ to the environment as it is cooled from $T_c$ to $T_1$ in the normal state, and
absorbs heat $Q_1$ from the environment as it is heated from $T_1$ to $T_c$ in the 
superconducting state. In addition, the system releases latent heat $L$ to the environment as it goes from normal to superconducting
at applied field $H_c(T_1)$ at temperature $T_1$. We assume that all the heat exchanges with the environment happen
under conditions where the system and the environment are at the same temperature, so that no net entropy is
generated by the heat transfer process.

We assume a system of unit volume for simplicity, and we assume that the radius of the
sample $R$  is much larger than the London penetration depth $\lambda_L$ , so that we don't have to deal with small 
corrections of order $\lambda_L/R$.
The latent heat is given by 
\beq
L(T)=T[S_n(T)-S_s(T)]
\eeq
with $S_n$ and $S_s$ the entropies in the normal and superconducting states. The heat absorbed by the system from the
environment in process 4 is
\beq
Q_1=\int_{T_1}^{T_c}C_s(T)dT
\eeq
and the heat released by the system to the environment in process 2 is
\beq
Q_2=\int_{T_1}^{T_c}C_n(T)dT
\eeq
where $C_s$ and $C_n$ are the heat capacities in the superconducting and normal states respectively. Using that
\beq
C_s(T)=T\frac{\partial S_s(T)}{\partial T} \;   \;   \;  \;   \;   \;  ;  \;   \;   \;  \;   \;   \;  C_n(T)=T\frac{\partial S_n(T)}{\partial T}
\eeq
integration by parts leads to
\beq
Q_1-Q_2-L(T_1)=\int_{T_1}^{T_c} dT[S_n(T)-S_s(T)] .
\eeq
where we have used that $S_n(T_c)=S_s(T_c)$. Eq. (5) is the net heat absorbed by the system  from the environment  in the cycle.
 
By conservation of energy, since the system is back to its initial state after the cycle, this energy 
has to have been delivered as work by the system to its environment, namely to  the power source
creating the magnetic field. During the transition from normal to superconducting,
the power source is supplying a constant current I generating the magnetic field
$H_c(T_1)$. As the system creates its own magnetic field in direction opposite to the applied field,
a counter-emf is generated through Faraday's law
\beq
\epsilon = -\frac{d\phi}{dt}
\eeq
where $\phi$ is the flux through the solenoid. This emf acts in the same direction as the current I supplied by the 
power source, as a consequence the energy per unit time delivered to the power source is
\beq
P=\frac{dW}{dt}=\epsilon I
\eeq
and the total energy delivered to the power source after the transition is completed is
\beq
W=I\Delta \phi=I\mu_0 H_c(T_1)A
\eeq
with $A$ the cross-sectional area of the superconducting cylinder.
From Ampere's law, the current in the solenoid generating the field $H_c(T_1)$
is $I=H_c(T_1)L$ where $L$ is the length of the solenoid and superconducting sample.
The volume of the superconducting sample is $A\times L$, which we assume to be unity, hence the 
total energy delivered to the power source by the sample of unit volume becoming superconducting is
\beq
W=\mu_0 H_c(T_1)^2 .
\eeq
Half of this energy is the energy of the magnetic field that was originally occupying the volume of the sample
and is no longer there when the sample is superconducting. We will assume that the other half is energy delivered by the 
sample in the process of going from normal to superconducting, \bluee{and show that this is consistent with the laws of thermodynamics. Instead,
Nikulov would argue that it is heat absorbed from the environment, which leads to violation of the second law of thermodynamics,
as discussed in the next section}.

Assuming the former, by energy conservation
\beq
Q_1-Q_2-L(T_1) = \mu_0 H_c(T_1)^2/2 .
\eeq
The internal energy of the sample is given by
\beq
U=F+TS
\eeq
where $F$ is the free energy. In going from normal to superconducting, the sample released latent heat $L(T_1)$ and
delivers work Eq. (10) to the power source. Therefore,
\beqn
U_n-U_s&=&\mu_0 H_c(T_1)^2/2+L(T_1)  \\ \nonumber
&=&F_n-F_s+T_1(S_n(T_1)-S_s(T_1))
\eeqn
Since the last term in this equation is the latent heat at temperature $T_1$, this implies that
\beq
F_n(T_1)-F_s(T_1)=\mu_0 H_c(T_1)^2/2 .
\eeq
From Eqs. (5) and (10) it follows that
\beq
\int_{T_1}^{T_c} dT[S_n(T)-S_s(T)] = \mu_0 H_c(T_1)^2/2
\eeq
and indeed since
\beq
S_n=-\frac{\partial F_n}{\partial T}  \;   \;   \;  \;   \;   \;  ;  \;   \;   \;  \;   \;   \;  S_s=-\frac{\partial F_s}{\partial T} 
\eeq
and $H_c(T_c)=0$ Eq. (14) follows from  Eqs. (13) and (15). Therefore, the Gorter cycle satisfies energy conservation, the
first law of thermodynamics, as expected.

\bluee{The reader may wonder how Eq. (13) is consistent with the expectation that in a first order 
phase transition at coexistence the free energies of the two phases should be equal. That is 
simply a matter of definition of free energies, as explained e.g. by Tinkham \cite{tinkham}. The Helmholtz free energy defined by 
Eq. (11) is not the same for the two phases because it does not include the energy of the current source that keeps the external
magnetic field constant, while an appropriately defined
Gibbs free energy is \cite{tinkham}. } 

Regarding entropy, the system has obviously the same entropy before and after the cycle is completed, since entropy is
a function of state. The environment decreased its entropy while releasing heat $Q_1$ to the sample, and
increased its entropy by absorbing heats $Q_2$ and $L$ from the sample. The net increase in entropy of the
environment in the cycle is
\beq
\Delta S_{env}=\int_{T_1}^{T_c} dT[\frac{C_n(T)}{T}-\frac{C_s(T)}{T}]+\frac{L(T_1)}{T_1}
\eeq
where the first term in the integral is the increase in entropy through absorption of $Q_2$ and the second term in the
integral is the decrease in entropy through release of heat $Q_1$. 
Using Eq. (4),
\beq
\Delta S_{env}=\int_{T_1}^{T_c} dT   [\frac{\partial S_n(T)}{\partial T} - \frac{\partial S_s(T)}{\partial T} ]            +\frac{L(T_1)}{T_1}
\eeq
and since $S_n(T_c)=S_s(T_c)$
\beq
\Delta S_{env}=-[S_n(T_1)-S_s(T_1)]           +\frac{L(T_1)}{T_1}  
\eeq
hence from Eq. (1) $\Delta S_{env}=0$.
Therefore, neither the environment nor the sample changed their entropy in the cycle, consistent with the fact that
it was a reversible process. The second law of thermodynamics is satisfied, not violated,
contrary to the claim in Ref. \cite{nikulov}.

The process discussed here is completely analoguous to a Carnot engine, where the system 
absorbs heat $Q_1$ from a heat reservoir  at higher temperature, releases heat $Q_2$ to a heat reservoir at lower
temperature, and delivers work $W=Q_1-Q_2$. 
The only difference is that in the Carnot cycle only two heat reservoirs at two different temperatures are needed,
while in the Gorter cycle an infinite number of reservoirs at temperatures in the interval
$T_1\leq T \leq T_c$ are needed.

Nikulov in Ref. \cite{nikulov} argues that
{\it ``The success of the conventional theory of
superconductivity forces us to consider the validity of belief in the law of entropy increase''}.
This is indeed correct, as discussed in the next section. We will argue that  an alternative conclusion is that {\it ``the success of the law of entropy increase
forces us to consider the validity of belief in the conventional theory of superconductivity''}.

%
%
%
%

\section{Inconsistency with the conventional theory of superconductivity, and Nikulov's alternative viewpoint}

To understand the contradiction with the conventional theory of superconductivity \bluee{under the assumption that the thermodynamic treatment discussed in the previous section is valid}, it is useful to consider
the  cycle shown in Fig. 1 in reverse. Because it is a reversible process, no change in the entropy of the universe occurs
in the reverse cycle either. Now in the reverse cycle,
the system transitions from the superconducting state carrying a surface current $I$ and the magnetic field excluded to the normal
state carrying no current and the magnetic field in the interior. During this process, the current has to stop without 
generation of entropy. 

Assume  the system instead of a superconductor was a normal metal, 
with a surface current $I$ that was induced through very rapid increase of the current supplied by the solenoid to the value I that generates
magnetic field $H_c(T_1)$, rapid enough that the surface current circulates over a skin depth much smaller than the radius of the 
system. Subsequently, the power source that is supplying a constant current I delivers energy $\mu_0 H_c(T_1)^2$ 
(Eq. 9) as the magnetic field penetrates the metallic sample, 
working against the counter-emf now opposing the direction of the solenoid current.
Half of that energy creates the magnetic field in the interior. The other half is dissipated as Joule heat, since in the normal 
metal the current
stops through scattering and dissipation:
\beq
Q_{Joule}=\mu_0 H_c(T_1)^2/2
\eeq
for a sample of unit volume. 

We can see this clearly   by consider the electromagnetic energy. From Faraday's and Ampere's law it follows that
\beq
-\vec{\nabla}\cdot(\vec{E}\times\vec{H})=\vec{J}\cdot\vec{E}+\frac{\partial}{\partial t} (\frac{\mu_0H^2}{2})
\eeq
and integrating over space and time
\beq
  -\int dt (\vec{E}\times\vec{H})\cdot d\vec{S}=  \int dt d^3r \vec{J}\cdot\vec{E}   +  \int d^3r (\frac{\mu_0H^2}{2})
\eeq
where the spatial integral on the left is over the surface of the sample and the volume integral on the right is over
the volume of the sample.
The left-hand term is the influx of electromagnetic energy into the sample,
the first term on the right is the work done by the electric field on charges, and the second term on the right is the
change in electromagnetic energy inside the volume.
The Faraday electric field at the lateral 
surface of the cylinder is 
\beq
\vec{E}(R,t)=-\frac{1}{2\pi R}\frac{\partial \phi}{\partial t}   \hat{\theta}
\eeq
pointing clockwise as seen from the top, where $\phi$ is the magnetic flux through the
cross section of the cylinder. The left-hand side of Eq. (21) giving the inflow of electromagnetic energy through
the lateral surface of the cylinder is:
\beq
  -\int dt (\vec{E}\times\vec{H})\cdot d\vec{S}=\frac{1}{2\pi R} \phi \times (2\pi RL)H=V\mu_0 H^2
  \eeq
  with $V=\pi R^2 L$ the volume of the cylinder of radius $R$ and length $L$  and $\phi=\mu_0 (\pi R^2) H$
  the magnetic  flux through a cross section of the cylinder at the end of the process. Hence from Eqs. (21) and (23)
  \beq
  \mu_0 H^2=  \frac{1}{V}\int dt d^3r \vec{J}\cdot\vec{E} +\frac{\mu_0 H^2}{2}
  \eeq
  with the first term on the right-hand side of Eq.  (24) the total work done by electric fields on charges inside the
  volume during the process, that for the normal metal sample gets dissipated as   Joule heat in agreement with Eq. (19).
  
  Instead, for a superconducting sample becoming normal, the current  at a given time $t$ and position $\vec{r}$ is in general a sum of 
  normal and superconducting currents:
  \beq
  \vec{J}(\vec{r},t)=\vec{J}_n(\vec{r},t)+\vec{J}_s(\vec{r},t)
  \eeq
  where one or the other or both could be non-zero. For example, in a superconducting region close to the
  boundary to a normal region there will be both a supercurrent $\vec{J}_s$ and a normal current $\vec{J}_n$
  of thermally excited quasiparticles. Eq. (24) then is
  \beqn
 & &  \mu_0 H^2=  \frac{\mu_0 H^2}{2}  \nonumber \\
  &+&  \frac{1}{V}\int dt d^3r \vec{J}_n\cdot\vec{E}   
+   \frac{1}{V} \int dt d^3r \vec{J}_s\cdot\vec{E}. 
  \eeqn
 Eq. (26) says that half of the electromagnetic energy that flows in goes into creating the magnetic field, and the 
 other half is the energy transmitted to normal and superconducting currents, both of which flow in direction parallel to
 the  Faraday electric field. Now from Eq. (13) we know that the system requires all that energy to transition
 into the normal state, hence we conclude that when the transition happens reversibly, which implies infinitely slowly,
 \bmath
 \beq
  \frac{1}{V}\int dt d^3r \vec{J}_n\cdot\vec{E}  =0
  \eeq
  \beq
   \frac{1}{V} \int dt d^3r \vec{J}_s\cdot\vec{E}= \frac{\mu_0 H^2}{2} = F_n-F_s 
  \eeq
  \emath
 and NO energy is dissipated as Joule heat.
 The question then arises: {\it how does the initial surface current plus the volume currents induced by the
 Faraday electric field during the transition, that carry total kinetic energy given by Eq. (27b), stop without generating
 Joule heat?}

It should  also be remembered that the supercurrent carries mechanical momentum \cite{momentum}. 
When the supercurrent stops its mechanical momentum cannot disappear, it has to be  transferred to the body as a whole. For  the normal metal where the current stops through scattering processes
that dissipate Joule heat the mechanical momentum is transferred to the body through these scattering processes. How is momentum transferred
to the body as a whole by the superconducting electrons when the supercurrent stops in the absence of collisions that would  involve scattering and dissipation?

It should also be remembered  that the fact that NO Joule heat  is generated in the process where the superconductor in a magnetic field
makes the transition to the normal state has been experimentally verified through numerous experiments by
W. H. Keesom and coworkers \cite{keesom}. The energy in question, Eq. (19), is proportional to the $volume$ of
the sample, so it would be easily detectable as Joule heat if it existed.

The conventional theory    \bluee{has not answered these questions, and
we argue that it cannot answer them because} it has no mechanism to transfer momentum
between electrons and the body as a whole without dissipation. Instead, the alternative theory
of hole superconductivity can \cite{book,holesc,holescreview}. The essential point is that holes are electrons with negative effective mass.
Electrons with negative effective mass interchange momentum with the body as a whole without dissipation
and perform the necessary momentum transfer.
How this happens in detail is discussed in  references \cite{momentum,disapp,entropy}.

\bluee{
Instead, according to Nikulov \cite{nikulov}, the supercurrent stops through development of Joule heat $Q_{Joule}$ in the
transition from the superconducting to the normal state. The same scattering processes transfer the momentum of the
supercurrent to the body as a whole, hence no other  mechanism to transmit momentum from the supercurrent to the
body as a whole without dissipation is required, consistent with the conventional theory of superconductivity that does not offer
such a mechanism. 
In the reverse process discussed in Sect. II, normal to superconductor transition, according to Nikulov the system absorbs heat $Q_{Joule}$ from the environment at the transition temperature
$T_1$, which generates the supercurrent and delivers energy to the power source, instead of obtaining that energy from the superconducting condensation energy. Eq. (10) gets replaced by
\beq
Q_1+Q_{Joule}-Q_2-L(T_1) = \mu_0 H_c(T_1)^2/2 .
\eeq
hence
\beq
Q_1-Q_2-L(T_1) =0
\eeq
and Eq. (13) gets replaced by
\beq
F_n=F_s
\eeq
and energy is conserved. The change in entropy of the universe for the cycle in Fig. 1 ran in reverse ($S\rightarrow N$ transition) is
\beq
\Delta S_{univ}= \frac{\mu_0 H_c(T_1)^2}{2T_1}
\eeq
and when the cycle is run as shown in Fig. 1 ($N\rightarrow S$ transition), the change in entropy of the universe is 
\beq
\Delta S_{univ}=-\frac{\mu_0 H_c(T_1)^2}{2T_1}
\eeq
in violation of the second law of thermodynamics. The cycle shown in Fig. 1 is a perpetuum mobile of the second kind
according to Nikulov's interpretation \cite{nikulov}.
}

\section{Conclusion}
\bluee{For completeness, I would also like to point out that the theory of hole superconductivity also provides
an explanation for how the momentum of the supercurrent develops in the N-S transition in a magnetic field and how this momentum
is compensated so that angular momentum conservation is not violated  \cite{pippard2015}, 
and we have argued that the conventional theory of superconductivity cannot explain it \cite{pippard2015}.
Nikulov instead  has argued that the supercurrent is propelled by an azimuthal ``quantum force'' \cite{quantumforce} but 
has not explained how this ``quantum force'' would be  consistent with momentum conservation.}

 I would like to also point out that in Ref. \cite{nikulov} Nikulov also argues 
 that observation of dc voltages in asymmetric superconducting  rings \cite{rings}  is another example of
 violation of the second law of thermodynamics by superconductors. 
 Whether  the theory of hole superconductivity can explain such observations
consistent with  thermodynamics is an open question.

In conclusion, I argue that we have shown here  that one of the following two alternatives has to be valid: (1) The Meissner effect violates the second law of thermodynamics, and is consistent with the BCS theory of superconductivity, as argued
by Nikulov  in \cite{nikulov}. (2) The Meissner effect is consistent with the second law of thermodynamics, 
establishes the invalidity of the BCS theory of superconductivity \cite{validity} \bluee{at least in its present form}, and is consistent with the theory
of hole superconductivity \cite{book,holesc,holescreview}. 

It is up to the reader to decide which of the two alternatives is more likely to be correct, \bluee{or to show that there is
a third alternative}.

\vspace{6pt} 

 \acknowledgments{I would like to express my  deep appreciation to A. V. Nikulov for   extensive discussions on these topics 
that greatly contributed to my understanding. }

\vspace{6pt}




 

 \end{document}